\documentclass{ws-procs10x7square}

\newcommand{\monh}{\mathbb{H}}
\newcommand{\monht}{\mathbb{H}_t}

\newcommand{\mony}{f}
\newcommand{\monk}{k}

\newcommand{\monK}{\mathbb{K}}
\newcommand{\monu}{\mathbb{U}}

\newcommand{\tild}{\widetilde}
\newcommand{\ket}[1]{| #1 \rangle}
\newcommand{\bra}[1]{\langle #1 |}
\newcommand{\aver}[1]{\langle \omega | #1 | \omega \rangle}
\newcommand{\statav}[1]{\langle #1 \rangle}

\begin{document}


\title{Conformal field theories in random domains\\ 
and Stochastic Loewner evolutions.}

\author{Denis Bernard}
\address{Service de Physique Th\'eorique de Saclay,
CEA/DSM/SPhT, Unit\'e de recherche associ\'ee au CNRS,
CEA-Saclay, 91191 Gif-sur-Yvette, email:dbernard@spht.saclay.cea.fr}  



\maketitle

\abstracts{We review the recently developed relation  between the
traditionnal algebraic approach to conformal field theories and
the more recent probabilistic approach based on stochastic Loewner
evolutions. It is based on implementing random conformal maps in
conformal field theories.}


\section{Introduction.}

Fractal critical clusters are hallmarks of criticality, 
as it may be illustrated by considering the $Q$-state Potts models
whose lattice partition functions are:
$$
Z= \sum_{\{s({\bf r})\}}\ \exp[\, J\sum_{{\bf r}\sqcup {\bf r'}}
\delta_{s({\bf r}),s({\bf r'})}\,]
$$
The sum is over all spin configurations and 
${\bf r}\sqcup {\bf r'}$ refers to neighbor sites ${\bf r}$ and
${\bf r'}$ on the lattice. The spin $s({\bf r})$ 
takes $Q$ possible values. 
By expanding the exponential factor 
using $\exp[J(\delta_{s({\bf r}),s({\bf r'})}-1)]=(1-p) +
p\delta_{s({\bf r}),s({\bf r'})}$ with $p=1-e^{-J}$,
these partition functions may be rewritten 
as sums over cluster configurations
$$
Z=e^{JL}\ \sum_C\, p^{\|C\|}\,(1-p)^{L-\|C\|}\, Q^{N_C}
$$
where $L$ is the number of links of the lattice, $N_C$  the number
of clusters in the configuration $C$ and 
$\|C\|$ the number of links inside the $N_C$ clusters, usually
called FK-clusters. Criticality is then encoded in 
the fractal nature of these clusters.

The stochastic Loewner evolutions (SLE) \cite{schramm} are
mathematically well-defined processes describing the growth of random
sets, called the SLE hulls, and of random curves, called the SLE
traces, embedded in the two-dimensional plane. The growths of these
sets are encoded into families of random conformal maps
satisfying specific evolution equations. Their distribution depends on 
a real parameter $\kappa$.

\begin{figure}[htbp]
  \begin{center}
    \includegraphics[width=0.4\textwidth]{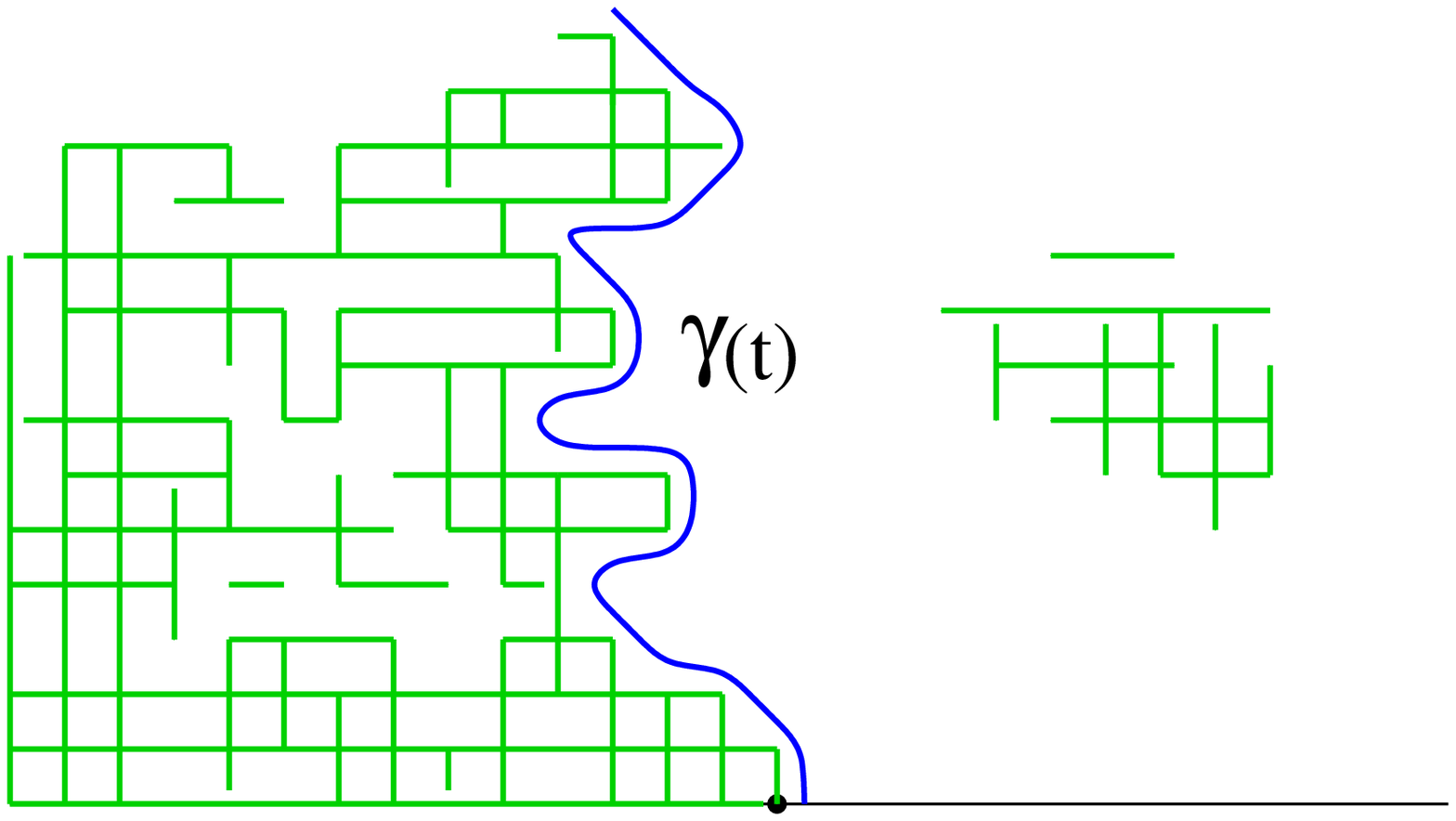}
      \caption{\em A FK-cluster configuration in the Potts model. The
        SLE trace is the boundary of the FK-cluster connected to
        the negative real line.}
      \label{fig:slepotts}
  \end{center}
\end{figure}

The connexion between critical systems and SLE growths may intuitively
be undertsood as follows: imagine considering the $Q$-state Potts
models on a lattice covering the upper half plane with boundary
conditions on the real line such that all spins on the negative real
axis are frozen to the same identical value while spins on the right
of the origin are free with non assigned values. Then, in each
configuration there exists a FK-cluster growing from the negative half
real axis into the upper half plane whose boundary starts at the
origin.  In the continuum limit, this boundary curve is conjectured to
be statistically equivalent to a SLE trace.  The SLE
parameter $\kappa$ is linked to the Potts parameter by
$Q=4\cos^2(\frac{4\pi}{\kappa})$, with $\kappa\geq 4$.  See Figure
(\ref{fig:slepotts}).

The aim of this note is to described a precise connexion, which we
have developed in refs.\cite{bb1,BBpartition,BBvir}, between the
 traditionnal algebraic approach to conformal field theories and
the probabilistic approach based on stochastic Loewner evolutions. 
The main point consists in considering conformal field theories on
random domains defined as the complements of the growing random SLE hulls.
Although we illustrate this connexion using the chordal version of
SLE,  we shall also touch upon the generalization to the radial SLE.

\section{(Chordal) stochastic Loewner evolutions.}

Given a simply connected domain $\mathbb{U}$ in the complex plane,
stochastic Loewner evolutions (SLE) describe the growth of random
curves emerging from the boundary of $\mathbb{U}$.  There are two
cases depending whether these curves connect two points on the
boundary $\partial \mathbb{U}$ (for the chordal SLE), or one point
on the boundary and one in the bulk of $\mathbb{U}$ (for the radial
SLE).  We shall mainly deal with the chordal case, except in the last
section.

To be more precise, let a hull in the upper half plane $\monh =\{z \in
\mathbb{C}, \Im {\rm m}\, z > 0\}$ be a bounded subset $\monK \subset \monh$
such that $\monh \setminus \monK$ is open, connected and simply
connected.

The local growth of a family of hulls $\monK_t$ parametrized by $t \in
[0,T[$ with $\monK_0=\emptyset$ is related to complex analysis as
follows. By the Riemann mapping theorem, $\monht\equiv \monh \setminus
\monK_t$, the complement of $\monK_t$ in $\monh$, which is simply
connected by hypothesis, is conformally equivalent to $\monh$ via a
map $\mony _t$.  This map can be normalized to behave as $\mony _t
(z)=z+2t/z+O(1/z^2)$, using the $PSL_2(\mathbb{R})$ automorphism group
of $\monh$. The crucial condition of \textit{local} growth leads to
the Loewner differential equation
\begin{eqnarray}
\partial_t \mony_t(z)=\frac{2}{\mony_t(z)-\xi_t}\ ,\quad \mony_{t=0}(z)=z
\label{loew}
\end{eqnarray}
with $\xi_t$ a real function. For fixed $z$, $\mony_t(z)$ is
well-defined up to the time $\tau_z$ for which
$\mony_{\tau_z}(z)=\xi_{\tau_z}$.  

(Chordal) stochastic Loewner evolutions is obtained \cite{schramm} by
choosing $\xi_t=\sqrt{\kappa}\, B_t$ with $B_t$ a normalized Brownian
motion and $\kappa$ a real positive parameter so that
$\mathbb{E}[\xi_t\,\xi_s]=\kappa\,{\rm min}(t,s)$.  The SLE hull is
reconstructed from $\mony_t$ by $\monK_t=\{z\in{\monh}:\ \tau_z\leq
t\}$ and the SLE trace $\gamma_{[0,t]}$ by
$\gamma(t)=\lim_{\epsilon\to0^+}f_t^{-1}(\xi_t+i\epsilon)$.  Basic
properties of the SLE hulls and SLE traces are described in
\cite{schramm,rhodeschra,LSW}.  In particular, $\gamma_{[0,t]}$ is
almost surely a curve.  It is non-self intersecting and it coincides
with $\monK_t$ for $0<\kappa\leq 4$, while for $4<\kappa<8$ it
possesses double-points and it does not coincide with $\monK_t$.

For establishing contact with conformal field theories (CFT), it is
useful to define $\monk _t(z) \equiv \mony _t(z) - \xi_t$ which
satisfies the stochastic differential equation
$$ 
d \monk _t = \frac{2dt}{\monk_t}-d\xi_t.
$$
The conditions at spatial infinity satisfied by $\monk _t$ imply
that its germ there belongs to the group $N_-$ of germs of holomorphic
functions at $\infty$ of the form $z+\sum_{m \leq -1} f_{m} z^{m+1}$.
The group $N_-$ acts on itself by composition,
$\gamma_f\cdot F\equiv F \circ f$ for $f \in N_-$, and
$\gamma_{g\circ f}= \gamma_f \cdot\gamma_g$.  In particular, to $k_t$
we can associate $\gamma_{k_t}\in N_-$, which satisfy, by It\^o's
formula:
$$
d \gamma_{\monk _t}\cdot F=(\gamma_{\monk _t}\cdot
F')(\frac{2dt}{\monk_t}-d\xi_t)+
\frac{\kappa}{2}(\gamma_{\monk _t}\cdot F'')
$$
Alternatively, this may be read as:
\begin{eqnarray}
\gamma_{\monk _t}^{-1} \cdot d \gamma_{\monk _t}=dt(\frac{2}{z}\partial_z
+\frac{\kappa}{2}\partial_z^2)-d\xi_t\partial_z.
\label{itoN-}
\end{eqnarray} 

The operators $l_n=-z^{n+1}\partial_z$ are represented in conformal
field theories by operators $L_n$ which satisfy the Virasoro algebra
$\mathfrak{vir}$~:
$$
[L_n,L_m] = (n-m)L_{n+m} +\frac{c}{12}(n^3-n) \delta_{n+m,0} \qquad
[c,L_n]=0.
$$

The representations of $\mathfrak{vir}$ are not automatically
representations of $N_-$, one of the reasons being that the Lie
algebra of $N_-$ contains infinite linear combinations of the $l_n$'s.
However, as we shall explain in the next section, highest weight
representations of $\mathfrak{vir}$ can be extended in such a way that
$N_-$ get embedded in a appropriate completion
$\overline{\mathcal{U}(\mathfrak{n}_{-})}$ of the enveloping algebra
of some subalgebra $\mathfrak{n}_{-}$ of $\mathfrak{vir}$.  This
will allows us to associate to any $\gamma_f\in N_-$ an operator $G_f$
acting on appropriate representations of $\mathfrak{vir}$ and
satisfying $G_{g \circ f}=G_f G_g$.  Implementing this construction to
$k_t$ yields random operators $G_{\monk _t}\in
\overline{\mathcal{U}(\mathfrak{n}_{-})}$ which satisfy the
stochastic It\^o equation:
\begin{eqnarray}
 G_{\monk _t}^{-1} d G_{\monk _t}=dt(-2L_{-2}+\frac{\kappa}{2}
L_{-1}^2)+d\xi_t L_{-1}.
\label{labelle}
\end{eqnarray}  
Compare with eq.(\ref{itoN-}).
This may be viewed as defining a Markov process in 
$\overline{\mathcal{U}(\mathfrak{n}_{-})}$.

Since $G_{\monk_t}$ turn out to be the operators intertwining the
conformal field theories in $\mathbb{H}$ and in the random domains
$\mathbb{H}_t$, the basic observation which allows us to couple CFTs
to SLEs is the following \cite{bb1}:

\vspace{.3cm}
\noindent {\it 
Let $\ket{\omega}$ be the highest weight vector in the irreducible
highest weight representation of $\mathfrak{vir}$ of central charge
$c_{\kappa}=\frac{(6-\kappa)(3\kappa-8)}{2\kappa}$ and conformal
weight $h_{\kappa}\equiv h_{1;2}=\frac{6-\kappa}{2\kappa}$.\\
Then $\mathbb{E}[G_{\monk _t}\ket{\omega}|\{G_{\monk_u}\}_{u\leq s}]$
is time independent for $t\geq s$ and:} 
\begin{eqnarray}
\mathbb{E}[\,G_{\monk _t}\ket{\omega}\,|\{G_{\monk_u}\}_{u\leq s}]
=G_{\monk _s}\ket{\omega}
\label{maineq}
\end{eqnarray}

This result is a direct consequence of eq.(\ref{labelle}) and the null
vector relation $(-2L_{-2}+\frac{\kappa}{2}L_{-1}^2)\ket{\omega}=0$
so that $dG_{\monk _t}\ket{\omega}=G_{\monk _t}L_{-1}\ket{\omega}d\xi_t$.

This result has the following consequences.  Consider CFT correlation
functions in $\monht$. They can be computed by looking at the same
theory in $\monh$ modulo the insertion of an operator representing the
deformation from $\monh$ to $\monht$, see the next section.  This
operator is $G_{\monk_t}$. Suppose that the central charge is
$c_{\kappa}$ and the boundary conditions are such that there is a
boundary changing primary operator of weight $h_{\kappa}$ inserted at
the tip of $\monK_t$. Then in average the correlation functions of the
conformal field theory in the fluctuating geometry $\monht$ are time
independent and equal to their value at $t=0$.

\begin{figure}[htbp]
  \begin{center}
    \includegraphics[width=0.6\textwidth]{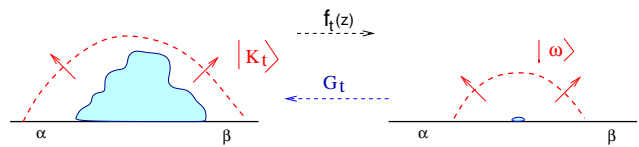}
      \caption{\em A representation of the boundary hull state and of the 
        map intertwining different formulations of the CFT.}
     \label{fig:hilbert}
 \end{center}
\end{figure}

The state $G_{\monk_t}\ket{\omega}$ may be interpreted as follows.
Imagine defining the conformal field theory in $\monh_t$ via a radial
quantization, so that the conformal Hilbert spaces are defined over
curves topologically equivalent to half circles around the origin. Then, 
the SLE hulls manifest themselves as disturbances localized around
the origin, and as such they generate states in the conformal Hilbert
spaces. Since, $G_{\monk_t}$ intertwines the CFT in $\monh$ and in
$\monh_t$, these states are $G_{\monk_t}\ket{\omega}$ with
$\ket{\omega}$ keeping track of the boundary conditions.
See Figure (\ref{fig:hilbert}).

\section{Conformal transformations in CFT and applications.}

The basic principles of conformal field theory state that
correlation functions in a domain $\monu$ are
known once they are known in a domain $\monu _0$ and an explicit conformal
map $f$ from $\monu$ to $\monu _0$ preserving boundary conditions is given.
Primary fields have a very simple behavior under conformal
transformations: for a bulk primary field $\varphi$ of weight
$(h,\overline{h})$,
$\varphi(z,\overline{z})dz^hd\overline{z}^{\overline{h}}$ is
invariant, and for a boundary conformal field $\psi$ of weight
$\delta$, $\psi(x)|dx|^{\delta}$ is invariant. Their statistical
correlations in $\monu $ and $\monu _0$ are related by
\begin{eqnarray}
 \statav{\cdots \varphi(z,\overline{z})\cdots \psi(x)\cdots}_{\monu} 
=\statav{\cdots \varphi(f(z),\overline{f(z)})f'(z)^h
  \overline{f'(z)}^{\overline{h}}\cdots
  \psi(f(x))|f'(x)|^{\delta}\cdots}_{\monu_0}. 
\label{cftconf}
\end{eqnarray}

Infinitesimal deformations of the underlying geometry are implemented
in local field theories by insertions of the stress-tensor.  In
conformal field theories, the stress-tensor is traceless so that it has
only two independent components, one of which, $T(z)$, is holomorphic
(except for possible singularities when its argument approaches the
arguments of other inserted operators).  The field $T(z)$ itself is not a
primary field but a projective connection,
$$\statav{\cdots T(z) \cdots}_\monu= \statav{\cdots
  T(f(z))f'(z)^2+\frac{c}{12}\mathrm{S}f(z)\cdots}_{\monu _0},$$
with $c$ is the CFT central charge and
$\mathrm{S}f(z)= \left(\frac{f''(z)}{f'(z)}\right)'-\frac{1}{2}
\left(\frac{f''(z)}{f'(z)}\right)^2$ 
the Schwartzian derivative of $f$ at $z$.

This applies to infinitesimal deformations of the upper
half plane. Consider an infinitesimal hull $\monK_{\epsilon;\mu}$, whose boundary 
is the curve $x\to\epsilon\,\pi\mu(x)$, $x$ real and $\epsilon\ll1$, 
so that $\monK_{\epsilon;\mu}=\{z=x+iy\in \monh,\ 0<y<\epsilon\,\pi\mu(x)\}$. 
Assume for simplicity that $\monK_{\epsilon;\mu}$ is 
bounded away from $0$ and $\infty$. 
Let $\monh_{\epsilon;\mu}\equiv \monh\setminus\monK_{\epsilon;\mu}$. 
To first order in $\epsilon$, its uniformizing map onto $\monh$ is 
$$
z+ \epsilon\,\int_{\mathbb{R}}\frac{\mu(y)dy}{z-y} + o(\epsilon).
$$ 
To first order in $\epsilon$,
correlation functions in $\monh_{\epsilon; \mu}$
 are related to those in $\monh$ by insertion of $T$:
\begin{eqnarray}
\frac{d}{d\epsilon}
\statav{\left(\cdots \varphi(z,\overline{z})\cdots
  \psi(x)\cdots\right)}_{\monh_{\epsilon; \mu}} \vert_{\epsilon=0^+} 
=\int_{\mathbb{R}} dy \mu(y)\, 
\statav{ T(y)\left(\cdots \varphi(z,\overline{z})\cdots
  \psi(x)\cdots\right)}_{\monh}
\label{Tdeform}
\end{eqnarray}
With the basic CFT relation \cite{bpz} between the stress tensor and
the Virasoro generators, $T(z)=\sum_n L_nz^{n-2}$, this indicates that
infinitesimal deformations of the domains are described by insertions
of elements of the Virasoro algebra.

Finite conformal transformations are implemented in conformal field
theories by insertion of operators, representing some appropriate
exponentiation of insertions of the stress tensor. Let
$\mathbb{U}$ be conformally equivalent to the upper half plane $\monh$
and $f$ the corresponding uniformazing map.  Then, following
\cite{BBpartition}, the finite conformal deformations that leads from
the conformal field theory on $\mathbb{U}$ to that on $\mathbb{H}$ can
be represented by an operator $G_f$:
\begin{eqnarray}
\statav{\cdots \varphi(z,\overline{z}) \cdots \psi(x) \cdots }_\mathbb{U}=
 \statav{ G_f ^{-1}\left(\cdots {\varphi}(z,\overline{z})\cdots 
  {\psi}(x) \cdots \right) G_f}_{\monh}.
\label{utoh}
\end{eqnarray}
This relates correlation
functions in $\mathbb{U}$ to correlation functions in $\mathbb{H}$ where the
field arguments are taken at the same point but conjugated by $G_f$.
Radial quantization is implicitely asssumed in eq.(\ref{utoh}).
Compare with eq.(\ref{cftconf}).

The following is a summary, extracting the main steps, of a
construction of $G_f$ described in details in \cite{BBpartition}.
To be more precise, we need to distinguish cases depending whether $f$
fixes the origin or the infinity.  We also need a few simple
definitions.  We let $\mathfrak{vir}$ be the Virasoro algebra
generated by the $L_n$ and $c$, and $\mathfrak{n}_{-}$ (resp.
$\mathfrak{n}_{+}$) be the nilpotent Lie subalgebra of
$\mathfrak{vir}$ generated by the $L_n$'s, $n <0$ (resp. $n >0$), and
by $\mathfrak{b}_{-}$ (resp.  $\mathfrak{b}_{+}$) the Borel Lie
subalgebra of $\mathfrak{vir}$ generated by the $L_n$'s, $n \leq 0$
(resp $n \geq 0$) and $c$. We denote by
$\overline{\mathcal{U}(\mathfrak{n}_{-})}$ 
(resp. $\overline{\mathcal{U}(\mathfrak{n}_{+})}$) appropriate
completion of the enveloping algebra of $\mathfrak{n}_{-}$
(resp. $\mathfrak{n}_{+}$). We shall only consider highest weight
vector representations of the Virasoro algebra.

\vskip .3 truecm

$\bullet$ {\it Finite deformations fixing $0$.}

Let $N_+$ be the space of power series of the form $z+\sum_{m \geq 1}
f_m z^{m+1}$ which have a non vanishing radius of convergence. With
words, $N_+$ is the set of germs of holomorphic functions at the
origin fixing the origin and whose derivative at the origin is $1$. In
applications to the chordal SLE, we shall only need the case when the
coefficients are real. But it is useful to consider the $f_m$'s
as independent commuting indeterminates.

$N_+$ is a group for composition.  Our aim is to construct a group
(anti)-isomorphism from $N_+$ with composition onto a subset
$\mathcal{N}_+ \subset \overline{\mathcal{U}(\mathfrak{n}_{+})}$ with
the associative algebra product.
 
We let $N_+$ act on $O_0$, the space of germs of holomorphic functions
at the origin, by $\gamma_f\cdot F\equiv F \circ f$ for $f \in N_+$
and $F \in O_0$. This representation is faithful and $\gamma_{g\circ
  f}=\gamma_{f} \gamma_{g}$. We need to know how $\gamma_f$ varies
when $f$ varies as $f\to f+\varepsilon v(f)$ for small $\varepsilon$
and an arbitrary vector field $v$.  Taking $g=z+\varepsilon v(z)$ in
the group law leads to $\gamma_{f+\varepsilon v(f)}F=\gamma_{f}\cdot F
+\varepsilon \gamma_{f}\cdot(v \cdot F) +o(\varepsilon)$, where $v
\cdot F(z) \equiv v(z)F'(z)$ is the standard action of vector fields
on functions.  Using Lagrange inversion formula to determine the
vector field $v$ corresponding to the variation of the indeterminate
$f_m$ yields:
$$
\gamma_f^{-1}\,\frac{\partial
  \gamma_f}{\partial f_m}= \sum_{n\geq m} 
\left(\oint_0 \frac{dw}{2i\pi} 
w^{m+1}\frac{f'(w)}{f(w)^{n+2}}\right)\, z^{n+1}\partial_z.
$$ 
This system of first order partial differential equations makes sense
in $\overline{\mathcal{U}(\mathfrak{n}_{+})}$ if we replace
$l_n=-z^{n+1}\partial_z$ by $L_n$. So, we define a connection $A_m$ in
$\overline{\mathcal{U}(\mathfrak{n}_{+})}$ by
\begin{eqnarray*}
A_m(f) \equiv  \sum_{n\geq m}
L_n\, \left(\oint_0 \frac{dw}{2i\pi} w^{m+1}\frac{f'(w)}{f(w)^{n+2}}\right)
 \label{eq:zerocurv}
\end{eqnarray*}
which by construction satisfies the zero curvature condition, 
$\frac{\partial A_l}{\partial f_k}-\frac{\partial A_k}{\partial
  f_l}=[A_k,A_l].$

We may thus construct an element $G_f \in \mathcal{N}_+\subset
\overline{\mathcal{U}(\mathfrak{n}_{+})}$ 
for each $f \in N_+$ by solving the system 
\begin{equation}
  \label{eq:pdesys}
\frac{\partial G_f}{\partial f_m}= -G_f A_m(f), \qquad m \geq 1.  
\end{equation}
This system is guarantied to be compatible, because $N_+$ is convex
and the representation of $N_+$ on $O_0$ is well defined for finite
deformations $f$, faithful and solves the analogous system.  The
existence and unicity of $G_f$, with the initial condition $G_{f=z}=1$,
is clear and the group (anti)-homomorphism property,
$G_fG_g=G_{g\circ f}$, is true because it is true infinitesimally and
$N_+$ is convex.  To lowest orders:
$G_f=1-f_1L_1+\frac{f_1^2}{2}(L_1^2+2L_2)-f_2L_2+\cdots.$

The element $G_{f}$, acting on a highest weight representation of the
Virasoro algebra, is the operator which implements the conformal map
$f$ in conformal field theory.  It acts on the stress tensor by
conjugaison as:
\begin{equation}
\label{eq:conjt}
G_f^{-1}T(z)G_f= T(f(z))f'(z)^2+\frac{c}{12}Sf(z).
\end{equation} 
A formula which makes sense as long as $z$ is in the disk of
convergence of $f(z)$ and $Sf(z)$, but which can be extended by
analytic continuation if $f(z)$ allows it.  A similar formula would
hold if we would have consider the action of $G_f$ on local fields. In
particular, by eq.(\ref{eq:conjt}), $G_f$ induces an homomorphism of the
Virasoro algebra by $L_m\to L_m(f)\equiv G_f^{-1}L_mG_f$ with
$G_f^{-1}T(z)G_f=\sum_mL_m(f)z^{-m-2}$:
\begin{eqnarray}
  \nonumber 
G_f^{-1}L_mG_f = \frac{c}{12}\,\left( 
\oint_0 \frac{dw}{2i\pi} w^{m+1}Sf(w)\right)
+\sum_{n\geq m}L_n\, \left( \oint_0 \frac{dw}{2i\pi} w^{m+1} 
\frac{f'(w)^2}{f(w)^{n+2}}\right).
\end{eqnarray}

Eq.(\ref{eq:pdesys}), which specifies the variations of $G_f$,
can be rewritten in a maybe more familiar way involving the stress
tensor. Namely, if $f$ is changed to $f+\delta f$ with $\delta
f=\varepsilon v(f)$, then:
$$
\delta G_f =- \varepsilon\, G_f\, \oint_0 \frac{dz}{2i\pi}\,T(z)\,v(z).
$$
If $v$ is not just a formal power series at the origin, but a convergent
one in a neighborhood of the origin, we can freely deform contours in
this formula, thus making contact with the infinitesimal deformations
considered in eq.(\ref{Tdeform}).
\vskip .3 truecm

$\bullet$ {\it Finite deformations fixing $\infty$.}

All the previous considerations could be extended to the case
in which the holomorphic functions fix $\infty$ instead of
$0$. Let $N_-$ be the space of power series of the form $z+\sum_{m
  \leq -1} f_{m} z^{m+1}$ which have a non vanishing radius of
convergence. We let it act on $O_{\infty}$, the space of germs of
holomorphic functions at infinity, by $\gamma_f\cdot F\equiv F \circ
f$. The adaptation of the previous computations shows that
$\gamma_f^{-1}\frac{\partial \gamma_f}{\partial f_m}= \sum_{n\leq m}
\left(\oint_{\infty} \frac{dw}{2i\pi} w^{m+1}\frac{f'(w)}{f(w)^{n+2}}\right)
z^{n+1}\partial_z.$ We transfer this relation to
$\overline{\mathcal{U}(\mathfrak{n}_{-})}$ to define an (anti)-isomorphism
from $N_-$ to $\mathcal{N}_-
\subset\overline{\mathcal{U}(\mathfrak{n}_{-})}$ mapping $f$ to $G_f$ such
that
$$\frac{\partial
  G_f}{\partial f_m}= -G_f \sum_{n\leq m} L_n\, 
\left( \oint_{\infty} \frac{dw}{2i\pi}
w^{m+1}\frac{f'(w)}{f(w)^{n+2}}\right), \qquad m \leq -1.$$ 
\vskip .3 truecm

$\bullet$ {\it Dilatations and translations.}

We have been dealing with deformations around $0$ and
$\infty$ that did not involve 
dilatation at the fixed point: $f'(0)$ or $f'(\infty)$ was unity.
To gain some flexibility we may also 
authorize dilatations, say at the origin. The operator
associated to a pure dilatation $z\to\lambda z$ is $\lambda^{-L_0}$. 
One can view a general $f$ fixing $0$ as the composition
$f(z)=f'(0)(z+\sum_m f_mz^{m+1})$ of a deformation at $0$ with
derivative $1$ at $0$ followed by a dilatation, so that
$G_f=G_{f/f'(0)}\, f'(0)^{-L_0}.$ 

We may also implement translations. Suppose that
$f(z)=f'(0)(z+\sum_m f_mz^{m+1})$ is a generic invertible germ of
holomorphic function fixing the origin. If $a$ is in
the interior of the disk of convergence of the power series expansion
of $f$ and $f'(a)\neq 0$, we may define a new germ $f_a(z)\equiv
f(a+z)-f(a)$ with the same properties. The operators $G_f$ and
$G_{f_a}$ implementing $f$ and $f_a$ are then related by  
$G_{f_a}=e^{-aL_{-1}}\,G_f\,e^{f(a)L_{-1}}.$

\vskip .3 truecm

\begin{figure}[htbp]
  \begin{center}
    \includegraphics[width=0.4\textwidth]{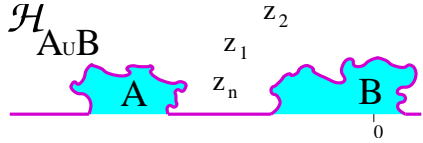}
      \caption{\em A typical two hull geometry.}
      \label{fig:sleblob}
  \end{center}
\end{figure}

$\bullet$ {\it Finite deformations around $0$ and $\infty$.}

Consider now a domain $\monh_{A\cup B}$ of the type represented on
fig.(\ref{fig:sleblob}) which is the complement two disjoint hulls, the
first one, say $A$, located around infinity but away from the origin,
and the second one, say B, located around the origin and away from infinity.
The uniformizing map $f_{A\cup B}$ of $\monh_{A\cup B}$ onto $\monh$ then
does not exist at $0$ or at $\infty$.

However, in this situation, we may obtain the map $f_{A\cup B}$ by
first removing $B$ by $f_B$, which is regular around $\infty$ and such
that $f_B(z)=z+O(1)$ at infinity, and then $\tild{A} \equiv f_B(A)$ by
$f_{\tild{A}}$ which is regular around $0$ and fixes $0$
($f_{\tild{A}}'(0)\neq 1$ is allowed).  
Of course, the roles of $A$ and $B$ could be interchanged, and we
could first remove $A$ by $f_A$ which is regular around $0$ and fixes
$0$ and then $\tild{B} \equiv f_A(B)$ by $f_{\tild{B}}$ which is
regular around $\infty$ and such that $f_{\tild{B}}(z)=z+O(1)$.

Suppose that $f_A$ and $f_B$ are given. There is some freedom in the
choice of $f_{\tild{A}}$ and $f_{\tild{B}}$~: namely we can replace
$f_{\tild{A}}$ by $h_0 \circ f_{\tild{A}}$ where $h_0\in
PSL_2(\mathbb{R})$ fixing $0$, and $f_{\tild{B}}$ by
$h_{\infty} \circ f_{\tild{B}}$ where $h_{\infty}\in
PSL_2(\mathbb{R})$ such that $h_{\infty}(z)=z+O(1)$ at 
infinity, i.e. a translation. A simple computation shows that
generically there is a unique choice of $f_{\tild{A}}$ and
$f_{\tild{B}}$ such that $f_{\tild{B}}\circ f_A=f_{\tild{A}}\circ
f_B$ and both equal to $f_{A\cup B}$.

For sufficiently disjoint hulls $A$ and $B$, as in
fig.(\ref{fig:sleblob}), there exists an open set such that for $z$ 
in this set, both
$G^{-1}_{f_{\tild{A}}}\left(G^{-1}_{f_B}T(z)G_{f_B}\right)G_{f_{\tild{A}}}$
and $G^{-1}_{f_{\tild{B}}}\left(G^{-1}_{f_A}T(z)G_{f_A}\right)G_{f_{\tild{B}}}$
are well defined, given by absolutely convergent series, and are both equal
to $T(f_{A\cup B}(z))f'_{A\cup B}(z)^2+\frac{c}{12}Sf_{A\cup B}(z)$.
As the modes $L_n$ of $T(z)$ generate all states in a highest weight
representation, the operators $G_{f_B}G_{f_{\tild{A}}}$ and
$G_{f_A}G_{f_{\tild{B}}}$ have to be proportional: they differ at
most by a factor involving the central charge $c$. We write
$G_{f_B}G_{f_{\tild{A}}}=Z(A,B)\; G_{f_A}G_{f_{\tild{B}}},$ or
\begin{equation}
  \label{eq:wik}
  G_{f_A}^{-1}\,G_{f_B}=Z(A,B)\, G_{f_{\tild{B}}}\,G_{f_{\tild{A}}}^{-1}.
\end{equation}

Formula (\ref{eq:wik}) plays for the Virasoro algebra the role that
Wick's theorem plays for collections of harmonic oscillators.
Since $G_{f_A}$ and $G_{f_{\tild{A}}}$ belong to $\mathcal{N}_+$ while 
$G_{f_B}$ and $G_{f_{\tild{B}}}$ to $\mathcal{N}_-$, eq.(\ref{eq:wik})
may also be viewed as defining a product between elements in
$\mathcal{N}_+$ and $\mathcal{N}_-$. Note that
$G_{f_{\tild{B}}}\,G_{f_{\tild{A}}}^{-1}$ is clearly well defined in
highest weight vector representations of $\mathfrak{vir}$.

As implicit in the notation, $Z(A,B)$ depends only on $A$ and $B$: a
simple computation shows that it is invariant if $f_A$ is replaced by
$h_0 \circ f_A$ and $f_B$ by $h_\infty \circ f_B$.  It may be
evaluated as follows. Let $A_s$ and $B_t$ be two families of hulls
that interpolate between the trivial hull and $A$ or $B$ respectively
and $f_{A_{s}}$ and $f_{B_{t}}$ be their uniformizing map. We arrange
that $f_{A_{s}}$ and $f_{B_{t}}$ satisfy the genericity condition, so
that unique $f_{A_{s,t}}$ and $f_{B_{t,s}}$ exist, which satisfy
$f_{B_{t,s}} \circ f_{A_s}=f_{A_{s,t}} \circ f_{B_{t}}$. Define vector
fields by $v_{A_{s}}$ and $v_{B_{t}}$ by $\frac{\partial
  f_{A_{s}}}{\partial s}=v_{A_{s}}(f_{A_{s}})$ and $\frac{\partial
  f_{B_{t}}}{\partial t}=v_{B_{t}}(f_{B_{t}})$. Then \cite{BBpartition}:
\begin{eqnarray}
\log Z(A_{\sigma},B_{\tau}) & = & \frac{c}{12}\int_0 ^{\sigma} ds
\oint \frac{dw}{2i\pi} \; v_{A_{s}}(w) Sf_{B_{\tau,s}}(w)\nonumber \\ 
& = & -\frac{c}{12}
\int_0 ^{\tau} dt \oint \frac{dz}{2i\pi} \; v_{B_{t}}(z)
Sf_{A_{\sigma,t}}(z).  
\label{parpart}
\end{eqnarray}
$Z(A,B)$ may physically be interpreted as the interacting part of the
CFT partition function in $\monh \setminus (A\cup B)$.

This two hull construction may be used to define operators which are
analogues of what vertex operators of dual models are for the
Heisenberg algebra. Indeed, consider a hull $A$ whose closure does
contain neither the origin nor the infinity.  Let us pick the
uniformizing map $f_A$ of its complement in $\mathbb{H}$ onto
$\mathbb{H}$ which is regular both at the origin and at infinity and
such that $f_A'(0)=f_A'(\infty)$.  Since it is regular at the origin,
we may implement $f_A$ in conformal field theory by
$G_{A^+}\,f_A'(0)^{-L_0}$ with $G_{A^+}$ in $\mathcal{N}_+$. Since it
is also regular at infinity, we may alternatively implement it by
$G_{A^-}\,f_A'(\infty)^{-L_0}$ with $G_{A^-}\in \mathcal{N}_-$.  The
product
$$
\mathcal{V}_A\equiv G_{A^-}\,G_{A^+}^{-1}
$$
is well defined and non trivial in highest weight
representation of $\mathfrak{vir}$. It may be thought of as the
factorization of the identity since the conformal transformation it
implements is the composition of two inverse conformal maps.

\section{Virasoro representations.}

The above formula may be used to define generalized coherent state
representations of $\mathfrak{vir}$. The key point is to interpret
the `Virasoro Wick theorem', eq.(\ref{eq:wik}), as defining an action
of $\mathfrak{vir}$ on $\mathcal{N}_-$. This is a reformulation of a
construction \`a la Borel-Weil presented in ref.\cite{BBvir}.
\vskip .3 truecm

$\bullet$ {\it Representations around infinity.}

Consider a Verma module $V(c,h)$ and take $x\neq 0$ its highest weight
vector. Let $f=z+\sum_{m \leq -1} f_m z^{m+1}\in N_-$
and $G_f$ be the corresponding element in $\mathcal{N}_-$.
The space $\{P_y[f]\equiv \left< y,G_fx \right>, y \in V(c,h)^*\}$, or
$\{Q_y[f]\equiv \langle y,G_f^{-1}x \rangle, y \in V(c,h)^*\}$,
is the space of all polynomials in the independent variables
$f_{-1},f_{-2},\cdots$.  So we have two linear isomorphisms from $V(c,h)^*$
to $\mathbb{C}[f_{-1},f_{-2},\cdots]$ and we can use them
to transport the action of $\mathfrak{vir}$.  We denote by
$\mathcal{R}_n$ and $\mathcal{S}_n$ the differential operators such
that
$$ 
\left<L_ny, G_fx \right>=\mathcal{R}_n \left< y,G_fx \right>
\quad ,\quad 
\langle L_ny,G_f^{-1}x \rangle=
\mathcal{S}_n \langle y,G_f^{-1}x \rangle
$$
for $y \in V^*(c,h)$. By construction the operators $\mathcal{R}_n$ and
$\mathcal{S}_n$ are first order differential operators satisfying the
Virasoro algebra with non vanishing central charge. 

To be more precise, let us first consider $P_y[f]=\left< y,G_fx
\right>$ and $P_{L_ny}[f]=\mathcal{R}_nP_y[f]$.  
We have $\left< L_ny,G_fx\right>=
\left<y,L_{-n}G_fx\right>$. If $n>0$,  $L_{-n}\in \mathfrak{n}_-$,
and, by the group law, the product $L_{-n}G_f$
corresponds to the infinitesimal variation of $f$ generated by 
$l_n=-z^{1-n}\partial_z$, namely $\delta_nf(z)=-z^{1-n}\, f'(z).$
If $n\leq0$, $L_{-n}\in \mathfrak{b}_+$ so that we need to re-order the
product $L_{-n}G_f$ in such way that it corresponds to an action
of $\mathfrak{b}_+$ associated to a variation of $f$.
This may be done using the Virasoro Wick theorem,
$G^{-1}_\phi G_f = Z(\phi,f)\, G_{\tild f}G_{\tild \phi}^{-1}$,
eq.(\ref{eq:wik}), which follows from the commutative diagram
$\tild \phi \circ f = \tild f \circ \phi$.
This diagram  shows that $\phi\in N_+$ acts on $N_-$ by $f\to \tild f$.
For $\phi(z)=z+\varepsilon z^{1-n}$, we have $\tild \phi(z)=z+\varepsilon
\tild y(z)$, with $\tild y$ polynomial of degree $1-n$, and
$\tild f= f + \varepsilon\delta_n f$ with 
\begin{eqnarray}
\delta_nf(z) = -z^{1-n}\, f'(z) + \tild y(f(z))
\label{varsource}
\end{eqnarray}
where $\tild y$ is fixed by demanding that $\delta_nf(z)=o(z)$ at
infinity. Namely, $\widetilde y(w)=\sum_ky_kw^{k+1}$ with 
$y_k=\oint_\infty \frac{dz}{2i\pi}\frac{z^{1-n}}{f(z)^{k+2}}f'(z)^2$.
Eq.(\ref{varsource}) are infinitesimal conformal transformations in
the source space generated by $\ell_n=-z^{1-n}\partial_z$ preserving the
normalization at infinity.
For $\phi(z)=z+\varepsilon z^{1-n}$, we also have 
$G^{-1}_\phi=1+\varepsilon L_{-n}$ and $G_{\tild
  \phi}^{-1}=1+\varepsilon (G_f^{-1}L_{-n}G_f)_{\mathfrak{b}_+}$
with $(G_f^{-1}L_nG_f)_{\mathfrak{b}_+}=\sum_k y_kL_k$.
The partition function is $Z(\phi,f)=1+\varepsilon\zeta$ with
$\zeta=\frac{c}{12}\oint_\infty \frac{dz}{2i\pi} z^{1-n} Sf(z)$.
As a consequence, $\left< L_ny,G_fx\right>=(\zeta+h y_0)\left<y,G_fx\right> 
+ \frac{d}{d\varepsilon}\langle y,G_{\tild f}x\rangle_{\varepsilon=0}$.
We thus get a representation of $\mathfrak{vir}$ with \cite{BBpartition}: 
\begin{eqnarray*}
\mathcal{R}_n & = & -\sum_{m \leq 0} (m+1)f_m \frac{\partial}{\partial f_{m-n}},
  \qquad n \geq 1 \\
\mathcal{R}_0 & = & h-\sum_{m \leq -1} mf_m\frac{\partial}{\partial f_m}\\
\mathcal{R}_{-1} & = & -2f_{-1}h-\sum_{m \leq -1}\Big(mf_{m-1}\; \; 
  -\sum_{k+l=m-1}f_kf_l\; \; +\; \; 2f_{-1}f_{m}\Big)
\frac{\partial}{\partial f_m}\\
\mathcal{R}_{-2} & = & -cf_{-2}/2-h(4f_{-2}-3f_{-1}^2)
+ \sum_{m \leq -1} (4f_{-2}-3f_{-1}^2)f_m \frac{\partial}{\partial f_m}\\
 & &  
-\sum_{m \leq -1} \Big( (m-1)f_{m-2}
-\sum_{j+k+l=m-2}f_jf_kf_l+ 3f_{-1}\sum_{k+l=m-1}f_kf_l\; \; \Big)
\frac{\partial}{\partial f_m}
\end{eqnarray*}
All other $\mathcal{R}_n$ are generated from these ones.

A similar construction may be used to deal with 
$Q_y[f]=\langle y, G_f^{-1}x \rangle$ 
giving formul\ae\ for $Q_{L_ny}[f]$ as a first order differential
operator $\mathcal{S}_n$ acting on $Q_y[f]$.
Once again the key point is that eq.(\ref{eq:wik}) allows to induce an 
action of $\mathfrak{vir}$ on $\mathcal{N}_-$. The operators
$\mathcal{S}_n$ and $\mathcal{R}_n$ are of course related as one goes
from ones to the others by changing $f$ into its inverse.
As a consequence the variation $f\to \tild
f=f+\varepsilon \bar \delta_n f$ induced by $L_n$ is now:
\begin{eqnarray}
\bar \delta_n f(z)= f(z)^{1-n} - \widehat y(z)f'(z)
\label{vartarget}
\end{eqnarray}
where $\widehat y(z)$ is fixed by demanding that $ \bar \delta_n
f(z)=o(z)$, ie. $\widehat y(z)= (f(z)^{1-n}/f'(z))_+$.
Eq.(\ref{vartarget}) are infinitesimal conformal transformations in the
target space generated by $\bar l_n=f^{1-n}\partial_f$ preserving the
normalization at infinity.
In particular, $\mathcal{S}_1$ corresponds to the variation
$\delta_1f=1$ and $\mathcal{S}_2$ to $\delta_2f=1/f$:
\begin{eqnarray*}
\mathcal{S}_1 & = & \frac{\partial}{\partial f_{-1}} \\
\mathcal{S}_{2} & = & \sum_{m\leq -2} 
\left(\oint_{\infty} dz\frac{1}{f(z)z^{m+2}}\right)
\frac{\partial }{\partial f_m}.  
\end{eqnarray*}
The other operators $\mathcal{S}_n$ may easily be found,
and are explicitely given in ref.\cite{BBvir}.
\vskip .3 truecm

$\bullet$ {\it Representations around the origin.}

The presentation parallels quite closely the case of deformations
around $\infty$ so we shall not give all the details. 
Let $f=z+\sum_{m \geq 1} f_m z^{m+1}$ be an element of $N_+$.
Consider a Verma module $V(c,h)$ and take $x$ its highest
weight vector. The space $\{\left< G_fy,x \right>, y \in
V(c,h)^*\}$, or $\{\langle G_f^{-1}y,x \rangle, y \in V(c,h)^*\}$,
 is the space of all polynomials in the independent variables
$f_1,f_2,\cdots$.  So we again have two linear
isomorphisms from $V(c,h)^*$ to $\mathbb{C}[f_1,f_2,\cdots]$, 
and we can use them to transport the
action of $\mathfrak{vir}$. This yields differential operators 
$\mathcal{P}_n$ and $\mathcal{Q}_n$ in the indeterminates $f_m$ such that:
$$
\left< G_fL_ny,x\right>=\mathcal{P}_n \left< G_fy,x \right>
\quad,\quad
\langle G_f^{-1}L_ny,x \rangle
=\mathcal{Q}_n \langle G_f^{-1}y,x \rangle
$$
for $y \in V^*(c,h)$. By construction the operators $\mathcal{P}_n$,
and $\mathcal{Q}_n$, satisfy the Virasoro algebra with central charge
$c$. Their expressions are given in \cite{BBpartition}. It is
interesting to notice the operators $\mathcal{Q}_n$, $n\geq 0$, coincide
with those introduced in matrix models. However, the above construction
provides a representation of the complete Virasoro algebra, with
central charge, and not only of one of its Borel subalgebras.
\vskip .3 truecm

$\bullet$ {\it Applications to SLE.}

We are now in position to rephrase the main result, eq.(\ref{maineq}),
in this language. Let $\mathcal{R}_n$ and $\mathcal{S}_n$ be the
differential operators define above and consider $f=\monk _t$ the SLE
map. Its coefficients $f_{-1},f_{-2},\cdots$ are random (for instance
$f_{-1}$ is simply a Brownian motion of covariance $\kappa$).  Because
$\mathcal{S}_n$, $n> 0$, are the differential operators implementing
the variation $\delta_n f=f^{1-n}$, the stochastic Loewner evolution
(\ref{loew}) may be written in terms of the Virasoro generators
$\mathcal{S}_n$ acting on functions $\mathcal{F}[f]$ of the $f_m$:
$$
d\mathcal{F}[f]
=dt\, (2\mathcal{S}_{2}+\frac{\kappa}{2}\mathcal{S}_{1}^2)\mathcal{F}[f]
- d\xi_t\,\mathcal{S}_{1}\mathcal{F}[f]
$$

Consider now the Verma module $V(c_\kappa,h_\kappa)$, with
$c_\kappa=\frac{(6-\kappa)(3\kappa-8)}{2\kappa}$ and
$h_\kappa=\frac{6-\kappa}{2\kappa}$. It is not irreducible, since
$(-2L_{-2}+\frac{\kappa}{2}L_{-1}^2)x$ is a singular vector in
$V(c_\kappa,h_\kappa)$, annihilated by the $L_n$'s, $n \geq 1$, so
that it does not couple to any descendant of $x^*$, the dual of $x$.
The descendants of $x^*$ in $V^*(c_\kappa,h_\kappa)$ generate the
irreducible highest weight representation of weight
$(c_\kappa,h_\kappa)$.  If $y$ is a descendant of $x^*$,
$\left<y,G_f(-2L_{-2}+\frac{\kappa}{2}L_{-1}^2)x \right>=0$, 
or equivalently,
$$ (2\mathcal{S}_{2}+\frac{\kappa}{2}\mathcal{S}_{1}^2)\left<y, G_fx
\right>=0$$
since, as fonction of the $f_m$, $\left<y, G_fL_{-n}x\right>
=-\mathcal{S}_n\left<y, G_fx \right>$ for $n \geq 1$.

As a consequence, all the
polynomials in $f_{-1},f_{-2},\cdots$ obtained by acting repeatedly on
the polynomial $1$ with the $\mathcal{R}_m$'s (they build the
irreducible representation with highest weight $(c_\kappa,h_\kappa)$) are
annihilated by $2\mathcal{S}_{2}+\frac{\kappa}{2}\mathcal{S}_{1}^2$.
For generic $\kappa$ there is no other singular vector in
$V(c_\kappa,h_\kappa)$, and this leads to a satisfactory description
of the irreducible representation of highest weight
$(c_\kappa,h_\kappa)$: the representation space is given by the
kernel of an explicit differential operator acting on
$\mathbb{C}[f_{-1},f_{-2},\cdots]$, and the states are build by repeated
action of explicit differential operators (the $\mathcal{R}_m$'s) on
the highest weight state $1$.

So the above computation can be interpreted as follows:
the space of polynomials of the coefficients of the expansion
of $\monk _t$ at $\infty$ for SLE$_\kappa$ can be endowed with a
Virasoro module structure isomorphic to $V^*(c_{\kappa},h_{\kappa})$.
Within that space, the subspace of (polynomial) martingales is a submodule
isomorphic to the irreducible highest weight representation of weight
$(c_{\kappa},h_{\kappa})$.

\section{Martingales and crossing probabilities.}

Let us now go to other applications  to
SLEs. As already mentionned the basic point is eq.(\ref{maineq}) 
which says that $G_{\monk_t}\ket{\omega}$ is a local martingale.
\vskip .3 truecm

$\bullet$ {\it The partition function martingale.}

The simplest application \cite{BBpartition} consists in using 
results of the previous two hull construction in the case when $B$ is
the growing SLE hull $\monK_t$ and $A$ is another disjoint hull away from
$\monK_t$ and the infinity. Let $f_A$ be the uniformizing map of
$\monh\setminus A$ onto $\monh$ fixing the origin.

Since $G_{\monk_t}\ket{\omega}$ is a local martingale, so is
$M_A(t)\equiv\aver{G_{f_A}^{-1}G_{\monk _t}}$.

To compute it, we start from $f_A$ and $\monk_t$ 
to build a commutative diagram as in previous section,
with maps denoted by $f_{\tild{A}_t}$ and $\tild{\monk}_t$
uniformizing respectively $\monk_t(A)$ and $f_A(\monK_t)$ and satisfying
$\tild{\monk}_t \circ f_A=f_{\tild{A}_t} \circ \monk_t$. Then 
$$
\aver{G_{f_A}^{-1}G_{\monk _t}} =
  Z(A,\monK_t)\,\aver{G_{\tild{\monk}_t}\,G_{f_{\tild{A}_t}}^{-1}}
$$
$Z(A,\monK_t)$ may be computed using eq.(\ref{parpart}):
$\log Z(A,\monK_t)={-\frac{c}{6}\int_0^{t} d\tau Sf_{A_{\tau}}(0)}.$
We have
$\aver{G_{\tild{\monk}_t}\,G_{f_{\tild{A}_t}}^{-1}}=
f_{\tild{A}_t}'(0)^{h_{\kappa}}.$
Thus the partition function martingale $M_A(t)$ reads:
$$M_A(t)=f_{\tild{A}_t}'(0)^{h_{\kappa}}\ \exp {-\frac{c_\kappa}{6}\int_0
  ^{t} d\tau\,  Sf_{A_{\tau}}(0)}.$$ 

This local martingale was discovered without any recourse to
representation theory in \cite{LSW}. We hope to have convinced
the reader that it is deeply rooted in CFT. From it, one
may deduce \cite{LSW} the probability that for $\kappa=8/3$
the SLE trace $\gamma_{[0,\infty]}$ does not touch $A$:
$$ \mathbb{P}[\gamma_{[0,\infty]}\cap A=\emptyset]= f_{A}'(0)^{5/8}$$
where $f_A$ has been further normalized by $f_A(0)=0$ and
$f_A(z)=z+O(1)$ at infinity.
Recall that for $\kappa=8/3\leq 4$, the SLE hull $\monk_t$ coincides
with the SLE trace $\gamma_{[0,t]}$ and that it almost surely avoids
the real axis at any finite time.
\vskip .3 truecm

$\bullet$ {\it Crossing probabilities.}

Crossing probabilities are probabilities associated to some stopping
time events. The approach we have been developing \cite{bb1} related
them to CFT correlations. It consists in projecting, in an appropriate way
depending on the problem, the martingale equation,  eq.(\ref{maineq}), which, 
as is well known, may be extended to stopping times. 
Given an event $\mathcal{E}$ associated to a stopping
time $\tau$, we shall identify a vector $\bra{v_\mathcal{E}}$ such
that $$\bra{v_\mathcal{E}}\, G_{\monk_\tau}\ket{\omega}= {\bf 1}_\mathcal{E}.$$
The martingale property of $G_{\monk_t}\ket{\omega}$ then
implies a simple formula for the probabilities:
$$
\mathbb{P}[\mathcal{E}]\equiv \mathbb{E}[\, {\bf 1}_\mathcal{E}\,] = 
\langle{v_\mathcal{E}}\ket{\omega}.
$$

For most of the considered events $\mathcal{E}$, the vectors
$\bra{v_\mathcal{E}}$ are constructed using conformal fields.  The
fact that these vectors  satisfy the appropriate
requirements, $\bra{v_\mathcal{E}}\, G_{\monk_\tau}\ket{\omega}= {\bf
  1}_\mathcal{E}$, is then linked to operator product expansion
properties \cite{bpz} of conformal fields.  This leads to express the
crossing probabilities in terms of correlation functions of 
conformal field theories defined over the upper half plane.
 
Consider for instance Cardy's crossing probabilities \cite{cardy}.
The problem may be formulated as follows.  Let $a$ and $b$ be two
points at finite distance on the real axis with
$a<0<b$ and define stopping times $\tau_a$ and
$\tau_b$ as the first times at which the SLE trace
$\gamma_{[0,t]}$ touches the interval $(-\infty,a]$ and $[b,+\infty)$.
The generalized Cardy's probability is the probability that
the SLE trace hits first the interval $(-\infty,a]$, that is
$\mathbb{P}[\,\tau_a<\tau_b\,]$. For this event, the vector
$\bra{v_\mathcal{E}}$ is constructed using the product of two
boundary conformal field $\psi_0(a)$ and $\psi_0(b)$ each of conformal 
weight $0$. This leads to the formula for $4<\kappa<8$:
$$
\mathbb{P}[\,\tau_a<\tau_b\,]=
\frac{\Phi_0(a/b)-\Phi_0(\infty)}{\Phi_0(0)-\Phi_0(\infty)}
$$
where $\Phi_0$ is the CFT correlation function, 
which only depends on $a/b$:
$$ \Phi_0(a/b)=\bra{\omega}\psi_0(a)\psi_0(b)\ket{\omega}.$$
More detailed examples have been described in \cite{bb1}.

Our approach and that of refs.\cite{LSW,schramm} are linked but they
are in a way reversed one from the other.  Indeed, the latter evaluate
the crossing probabilities using the differential equations they
satisfy -- because they are associated to martingales,-- while we
compute them by identifying them with CFT correlation functions --
because they are associated to martingales -- and as such they satisfy
the differential equations.

\section{(Radial) SLEs.}
We now briefly illustrate how previous results can be adapted to
deal with the radial stochastic Loewner evolutions. 
For CFT convenience, we prefer to view them as describing 
hulls $\breve{\mathbb{K}}_t$ growing outside
the unit disc centered at the origin, and not into as usual. 
Let $\mathbb{D}_x$ be the disc of unit radius centered
in $x$ and $\breve{\mathbb{D}}_x\equiv \mathbb{C} \setminus \mathbb{D}_x$ 
be its complement in the complex plane. The Loewner equation for the
radial SLE conformal map $g_t$ is:
\begin{eqnarray}
\partial_t g_t(z)=- g_t(z)\frac{g_t(z)+U_t}{g_t(z)-U_t}\quad,\quad
g_{t=0}=z \label{eq:radial}
\end{eqnarray}
with $U_t=e^{i\xi_t}$, a Brownian motion on the unit circle.
As for the chordal case, the SLE hulls are the set of points which
have been swallowed: $\breve{\mathbb{K}}_t=\{z\in
\breve{\mathbb{D}}_0; \tau_z\leq t\}$ with $\tau_z$ the swallowing
time such that $g_{\tau_z}(z)=U_{\tau_z}$. Since we view the hulls as
growing toward infinity, $g_t$ is the uniformizing map of the
complement of $\breve{\mathbb{K}}_t$ in $\breve{\mathbb{D}}_0$, and
it is normalized by $g_t(z)=e^{-t}z+O(1)$ at infinity.

For making contact with CFT, it is useful to translate the disc by
$-1$ so that the SLE hulls start to be created at the origin and
growth into $\breve{\mathbb{D}}_{-1}$.  So we define $h_t$ by
$h_t(z)+1=U^{-1}_tg_t(z+1)$.  Both $h_t$ and $g_t$ are regular at
infinity and, by the results of previous sections, we may associate to
them operators $G_{g_t}$ and $H_t\equiv G_{h_t}$ in 
$\overline{\mathcal{U}(\mathfrak{b}_-)}$
which implement these conformal maps in CFT.  They are linked by
$$H_t=e^{-L_{-1}}\, G_{g_t}\, e^{i\xi_tL_0}\, e^{L_{-1}}.$$
By It\^o calculus, $H_t$ satisfies the stochastic equation:
\begin{eqnarray}
H_t^{-1}dH_t = \left( -2w_{-2} +\frac{\kappa}{2}w_{-1}^2\right)\, dt 
+ w_{-1}\, d\xi_t 
\label{laradial}
\end{eqnarray}
with $w_{-1}=i(L_0+L_{-1})$ and 
$w_{-2}=-\frac{1}{2}(L_0+3L_{-1}+2L_{-2})$. These generators have a
simple interpretation: $w_{-1}$ generates rotations of
$\mathbb{D}_{-1}$ around its center, and $w_{-2}$ is the vector
generating infinitesimal slits at $0$ and away from
$\mathbb{D}_{-1}$.

As in the chordal case, a key remark is the following:
\medskip

\noindent { \it 
    Let $\ket{\omega}$ be the highest weight vector in the irreducible
    highest weight representation of $\mathfrak{vir}$ of central
    charge $c_{\kappa}=\frac{(6-\kappa)(3\kappa-8)}{2\kappa}$ and
    conformal weight $h_{\kappa}\equiv
    h_{1;2}=\frac{6-\kappa}{2\kappa}$.\\
    Let $d_\kappa\equiv 2h_{0;1/2}=\frac{(6-\kappa)(\kappa-2)}{8\kappa}$.\\
    Then $e^{-td_\kappa}\, H_t\ket{\omega}$ is a local martingale, in
    particular $\mathbb{E}[\,e^{-td_\kappa}\, H_t\ket{\omega}]$ is time
    independent.}
\medskip

This result follows from the fact that the evolution
operators $\mathcal{A}_r\equiv( -2w_{-2} +\frac{\kappa}{2}w_{-1}^2)$ reads
$$\mathcal{A}_r=(2L_{-2}-\frac{\kappa}{2}L_{-1}^2)+\kappa L_{-1}(h_\kappa-L_0)
+ L_0-\frac{\kappa}{2}L_0^2$$ so that 
$dH_t\ket{\omega}=d_\kappa H_t\ket{\omega}dt+
H_tw_{-1}\ket{\omega}d\xi_t$ with
$d_\kappa=h_\kappa-\frac{\kappa}{2}h_\kappa^2$. 
Notice that the radial evolution operators $\mathcal{A}_r$ has
a triangular structure contrary to the chordal one.

This may be used to construct the restriction martingale
coding for the influence of deformations of domains on SLE. Let $A$ be a
hull in $\breve{\mathbb{D}}_{-1}$ and $\phi_A$ be one of its
uniformizing map onto $\breve{\mathbb{D}}_{-1}$ fixing the origin,
$\phi_A(0)=0$.  Given $\phi_A$ and $h_t$, we may write in a unique way
a commutative diagram $\phi_{\tild A_t}\circ h_t = \tild h_t\circ
\phi_A$ where $\phi_{\tild A_t}$ (resp. $\tild h_t$) uniformizes
the complement of $h_t(A)$ (resp. $\phi_A(\breve{\mathbb{K}}_t)$) onto
$\breve{\mathbb{D}}_{-1}$ with $\phi_{\tild A_t}(0)=0$ and $\tild
h_t(\infty)=\infty$. Let $H_t$ (resp. $\tild H_t$) be the operator
implementing $h_t$ (resp. $\tild h_t$) in CFT. Similarly, let $G_{\phi_A}$
(resp. $G_{\phi_{\tild A_t}}$) be those implementing $\phi_A$
(resp.$\phi_{\tild A_t}$). 
Then, $G^{-1}_{\phi_A}\, H_t = Z_t(A)\, \tild
H_t\, G^{-1}_{\phi_{\tild A_t}}$ with
$$Z_t(A) =\exp \frac{c}{6} \int_0^tds\, 
S\phi_{\tild A_s}(0).$$

By construction $e^{-td_\kappa}\, G^{-1}_{\phi_A}\,H_t\ket{\omega}$ is a
local martingale. It is convenient to project it on the state
$\bra{\Omega}$ created by the bulk conformal operator of dimension
$2h_{0;1/2}$ located at infinity. (This is compatible with CFT fusion
rules.) Computing $\bra{\Omega} G^{-1}_{\phi_A}\,H_t\ket{\omega}$ using the
commutative diagram yields the martingale
$$ M_A^r(t) \equiv e^{-td_\kappa} |\tild h_t'(\infty)|^{-2h_{0;1/2}} 
\phi_{\tild A_t}'(0)^{h_\kappa}\, Z_t(A) $$
Alternatively, since $h_t'(\infty)\phi_{\tild A_t}'(\infty)=
\tild h_t'(\infty)\phi_A'(\infty)$ and $d_\kappa=2h_{0;1/2}$, this
reads:
$$ M_A^r(t)\, |\phi_{A}'(\infty)|^{-d_\kappa} 
=  \phi_{\tild A_t}'(0)^{h_\kappa}\,
|\phi_{\tild A_t}'(\infty)|^{-d_\kappa}\, Z_t(A)
$$
It may be further generalized by considering states created by bulk
operators of dimension $2h_{0;1/2}+\frac{\kappa}{2}s^2$ 
but of non trivial spin $s$. 
It may be used to evaluate the probability that the radial SLE hull at 
$\kappa=8/3$ does not touch $A$.

More details on the radial SLE will be described elsewhere
\cite{BBradial}.




\section*{Acknowledgments}
All results described in this note were obtained in collaboration with
Michel Bauer.

 Work supported in part by EC contract number
HPRN-CT-2002-00325 of the EUCLID research training network.











\begin{thebibliography}{0}


\bibitem{bpz} A. Belavin, A. Polyakov, A. Zamolodchikov,
  \textit{Infinite conformal symmetry in two-dimensional quantum field
    theory}, Nucl. Phys. {\bf B241}, 333-380, (1984).  

\bibitem{schramm} O. Schramm, Israel J. Math., {\bf 118}, 221-288,
  (2000);

\bibitem{rhodeschra} S. Rhode, O. Schramm, \textit{Basic properties of
    SLE}, and references therein, arXiv:math.PR/0106036.

\bibitem{LSW} G. Lawler, O. Schramm, W. Werner, \textit{Values of
    Brownian intersections exponents I : half-plane exponents}, Acta
  Mathematica {\bf 187} (2001) 237-273, arXiv:math.PR/9911084; \\
  G. Lawler, O. Schramm, W. Werner, \textit{Values of Brownian
    intersections exponents II : plane exponents}, Acta
  Mathematica {\bf 187} (2001) 275-308, arXiv:math.PR/0003156; \\
  G. Lawler, O. Schramm, W. Werner, \textit{Values of Brownian
    intersections exponents III : two-sided exponents}, Ann. Inst.
  Henri Poincar\'e {\bf 38} (2002) 109-123, arXiv:
  math.PR/0005294; \\
  G. Lawler, O. Schramm, W. Werner,\textit{ Conformal restriction: the
    chordal case}, arXiv:math.PR/0209343.

\bibitem{bb1} M. Bauer, D. Bernard, \textit{Conformal field theories
    of stochastic Loewner evolutions}, arXiv:hep-th/0210015, to appear
  in Commun. Math. Phys.

\bibitem{BBpartition} M. Bauer, D. Bernard, \textit{Conformal
    transformations and the SLE partition function martingale},
  arXiv:math-ph/0305061. 

\bibitem{BBvir} M. Bauer, D. Bernard \textit{$SLE$ martingales and
    the Virasoro algebra}, arXiv:hep-th/0301064, Phys. Lett. {\bf B557}
  (2003) 309-316.

\bibitem{smirnov} S. Smirnov, \textit{Critical percolation in the
    plane: conformal invariance, Cardy's formula, scaling limits},
  C.R. Acad. Sci. Paris, (2001) {\bf 333} 239-244.

\bibitem{cardy} J. Cardy, {\it Critical percolation in finite
    geometry}, J. Phys. {\bf A25}, L201-206, (1992).

\bibitem{BBradial} M. Bauer, D. Bernard, {\it CFTs of SLEs: the radial 
    case}, in preparation.

\end{thebibliography}
\end{document}